\documentclass[twocolumn,aps,prd,preprintnumbers]{revtex4}

\usepackage{graphicx}
\usepackage{dcolumn}
\usepackage{bm}
\usepackage{subfigure}
\usepackage{slashed}
\usepackage{marvosym}
\usepackage{textcomp}
\usepackage{threeparttable}
\usepackage{booktabs}
\usepackage{setspace}
\usepackage{latexsym}
\usepackage{mciteplus}
\renewcommand{\TPTtagStyle}%
{\normalsize\textit}

\begin{document}

{\raggedleft{}CCTP-2015-08, CCQCN -2015-79}
\title{Viscous Leptons in the Quark Gluon Plasma}
\author{Berndt M\"uller$^{1,2}$\footnote{muller@phy.duke.edu}, Di-Lun Yang$^3$\footnote{dy29@phy.duke.edu}}
\affiliation{
$^1$Department of Physics, Duke University, Durham, North Carolina 27708, USA.\\
$^2$Brookhaven National Laboratory, Upton, NY 11973, USA.\\
$^3$Crete Center for Theoretical Physics, Department of Physics
University of Crete, 71003 Heraklion, Greece.} %
\date{\today}
\begin{abstract}
We investigate the shear viscosity of leptons in a strongly coupled quark gluon plasma (QGP). We find that the shear viscosity due to the lepton-quark scattering is inversely proportional to the ratio of electric conductivity of the QGP to temperature up to the leading logarithmic order of the electromagnetic coupling. The finding implies that the thermal leptons form a more viscous fluid than the quarks by a large ratio. Using the known result for the electrical conductivity of strongly coupled plasmas obtained from gauge/gravity duality, we find that the lepton shear viscosity is suppressed compared with the one from lepton-lepton scattering. Consistently, we find an enhancement of the energy loss of hard leptons in a strongly coupled scenario compared with that in a weakly coupled plasma. 
\end{abstract}

\maketitle
The transport properties of thermal plasmas have been widely investigated for decades, because such plasmas play important roles in many areas of physics, ranging from thermonuclear fusion to astrophysics. Most recently, plasmas including free quarks have been studied in the context of relativistic heavy ion collisions and cosmology.  In weakly coupled scenarios described by quantum electrodynamics (QED) or perturbative quantum chromodynamics (QCD), transport coefficients such as the shear viscosity $\eta$ and electric conductivity $\sigma$ can be computed in the framework of kinetic theory \cite{Heiselberg:1994vy,Arnold:2000dr,Arnold:2003zc,Chen:2010xk}. Similar studies have been also carried out for weakly coupled hadronic gas and a variety of quantum field theories \cite{Jeon:1994if,Jeon:1995zm,Chen:2006iga,Chen:2007jq,Chen:2007xe,Chen:2007kx}. In strongly coupled scenarios, the perturbative calculation and the quasi-particle description underlying kinetic theory become invalid. However, the transport coefficients can still be evaluated non-perturbatively using Kubo formulas. For example, Demir and Bass \cite{Demir:2008tr} used the Kubo formalism combined with a phenomenological transport model to compute the ratio of shear viscosity to entropy density ($\eta/s$) of a thermal hadronic gas(see relevant studies of hadronic matter in \cite{NoronhaHostler:2008ju,NoronhaHostler:2012ug}). The gauge/gravity duality can be employed to evaluate $\eta$ for strongly coupled plasmas via the Kubo formula \cite{Policastro:2001yc,Kovtun:2004de} giving rise to the well-known Kovtun-Son-Starinets (KSS) bound of $\eta/s$. 

Here we address a more subtle situation -- the mixture of both weakly coupled and strongly coupled sectors of the plasma -- which exists in the practical cases such as the quark gluon plasma present in the early universe. The plasma comprises both leptons and colored quanta (quark and gluons), where the interactions among leptons are weakly coupled but the interactions among quarks and gluons are strongly coupled. The two sectors are connected by electroweak scattering between leptons and quarks. Intuitively, one would expect a minor influence of the leptonic sector on the QCD sector, while the magnitude of the  inverse effect is less obvious. We address this question by computing $\eta$ of light thermal leptons embedded in the strongly coupled plasma. 

Because of their weak coupling, the Boltzmann approach can be applied to analyze the leptonic transport. There exist two collisional terms: one corresponds to lepton-lepton scattering and the other comes from lepton-quark scattering, which encodes the non-perturbative nature of the QCD sector. We denote the former contribution to the viscosity of the QED plasma by $\eta_{\text{QED}}$ and the latter by $\eta_{\text{mix}}$.  Since the collisional terms are additive in the linear Boltzmann equation, one could compute $\eta_{\text{mix}}$ and $\eta_{\text{QED}}$ separately. The complete shear viscosity of leptons 
\begin{equation}
\eta_c\approx \frac{\eta_{\text{mix}}\eta_{\text{QED}}}{\eta_{\text{mix}}+\eta_{\text{QED}}}
\end{equation}
will be always smaller than each individual contribution.           

\begin{figure}[t]
{\includegraphics[width=4.5cm,height=2cm,clip]{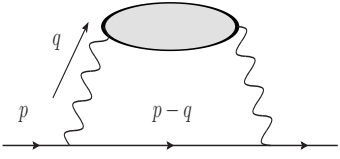}}
\caption{The self-energy diagram for thermal leptons. The curvy lines correspond to off-shell thermal photons and the solid lines represent the thermal leptons. The blob incorporates the strong coupling in the QCD sector to all orders.}\label{selfenergy_fig}
\end{figure}

We will now briefly describe our strategy to tackle the problem and mention our salient findings.
In the Boltzmann approach, the collisional term is proportional to the square of the scattering amplitude, which can be related to the imaginary part of the photon self energy through the optical theorem. The photon self energy can be further written as the current-current correlation function, which can be computed via perturbative or non-perturbative approaches. The schematic figure is illustrated in Fig.\ref{selfenergy_fig}. It was found in weak coupling that the collisional integral is dominated by an infrared (IR) divergence led by small momentum transfers. Such a divergence can be regularized in terms of a logarithmic power of the coupling, $\ln(1/e)$ for QED and $\ln(1/g)$ for pQCD, by introducing an IR cutoff around the scale of Debye mass \cite{Heiselberg:1994vy,Arnold:2000dr}, which gives the so-called leading logarithmic results of transport coefficients. 

In the strongly coupled scenario (in the QCD sector), we will use the correlation function obtained from the gauge/gravity duality in the large number of colors $N_c$ limit for the lepton-quark collision. Given that small-momentum transfers dominates\, we focus on the long-wavelength limit. It turns out that the collisional integral is mainly contributed by collinear divergence when the intermediate virtual photons become almost lightlike. By ignoring the contribution from the longitudinal part of the correlator, we further relate $\eta_{\text{mix}}$ to the electric conductivity $\sigma_c$ of the color sector. We find that 
\begin{equation}
\eta_{\text{mix}}/T^3\sim T/\sigma_c,
\end{equation}
which is physically intuitive since the strength of collisions between two sectors is characterized by $\sigma_c$ in the long-wavelength limit. When $\sigma_c/T$ is large, $\eta_{\text{mix}}$ will be suppressed making thermal leptons more viscous. 

We also study the energy loss for high-energy leptons or heavy leptons like muons, which slowly thermalize in the medium. We find that their scattering with the QCD sector results in a linear divergence in the IR regime for heavy leptons, which dominates over the logarithmic divergence from the scattering with thermal leptons. For light leptons, the $1/N_c$ suppression of $\eta_{\text{mix}}$ and the $\mathcal{O}(N_c)$ enhancement of the energy loss mutually support each other. In the following, we now present the details of our approach and derivations. We denote the electric charge of the leptons by $e$ and that of the single flavor of quarks by $Q$. The generalization to several quark flavors with differing electric charges is straightforward.

In thermal equilibrium, the interaction rate of leptons in the relativistic Boltzmann approach can be written as
\begin{eqnarray}\label{intrates}
\frac{p^{\mu}}{E_p}\partial_{\mu}\tilde{f}(p,x)=-\tilde{f}(p,x)\Gamma^{>}(p)+(1-\tilde{f}(p,x))\Gamma^{<}(p),
\end{eqnarray}     
where $\tilde{f}(p,x)$ denotes the distribution function of leptons in phase space. Here $\Gamma^{>}(p)$ and $\Gamma^{<}(p)$ represent the radiation and absorption rates, respectively. Based on the optical theorem, one finds 
\begin{eqnarray}\label{decay_rates}
\Gamma^{>(<)}(p)&=&(-)\frac{1}{4E_p}\textbf{tr}\left[(\slashed{p}+m)\Sigma^{>(<)}(p)\right],
\end{eqnarray}
where $m$ is the lepton mass. The $\Sigma^{>(<)}(p)$ are given by
\begin{eqnarray}
\Sigma^{>}(p)=e^2\int\frac{d^4q}{(2\pi)^4(q^2)^2}\gamma^{\mu}\Pi^{>}_{\mu\nu}(q)S_{F}^{>}(p-q)\gamma^{\nu},
\end{eqnarray}
and the analogous equation for $\Sigma^{<}(p)$, where $S_{F}^{>}$ and $\Pi^{>}_{\mu\nu}(q)$ are Wightman functions of on-shell leptons and electromagnetic currents generated by thermal quarks in momentum space. In thermal equilibrium, one may rewrite $\Pi^{>(<)}_{\mu\nu}(q)$ in terms of the retarded correlator $\Pi^R_{\mu\nu}(q)$ via linear response theory.  Note that $\Pi_{\mu\nu}^R(q)$ is proportional to $Q^2$ since we only consider the leading-order interaction for the electromagnetic coupling. However, $\Pi_{\mu\nu}^R(q)$ contains all orders of the strong coupling $g$. One could easily show that $\Gamma^{>}(p)=e^{\beta E_p}\Gamma^{<}(p)$ expressing detailed balance.  

The Boltzmann equation now becomes
\begin{eqnarray}\label{Boltzmann_eq}
p^{\mu}\partial_{\mu}\tilde{f}(p,x)&=&e^2\int\frac{d^3\bf{q}}{(2\pi)^3 4E_{p-q}}
\frac{\epsilon(q_0)\text{Im}[\Pi^R_{\mu\nu}(q)]}{(q^2)^2}
\\\nonumber
&&\text{tr}\left[(\slashed{p}+m)\gamma^{\mu}((\slashed{p}-\slashed{q})+m)\gamma^{\nu}\right]D(q, p),
\end{eqnarray}
where $E_{p-q}=\sqrt{|{\bf{p}-\bf{q}}|^2+m^2}$ and
\begin{eqnarray}\nonumber
D(q,p)&=&\tilde{f}(p,x)(1+f(q,x))(1-\tilde{f}(p-q,x))
\\
&&-(1-\tilde{f}(p,x))f(q,x)\tilde{f}(p-q,x).
\end{eqnarray}
Here $\tilde{f}(p,x)$ and $f(p,x)$ correspond to fermionic and bosonic distributions and $\epsilon(q_0)$ denotes the sign function of $q_0$ following the convention in \cite{Bellac:1996}. One can easily check that $D(q,p)=0$ when $\tilde{f}(p,x)$ and $f(p,x)$ are the thermal distribution functions. 

Following the general approach in \cite{Jeon:1994if,Jeon:1995zm,Dusling:2009df}, we introduce 
the perturbations away from equilibrium for the leptons as
$\tilde{f}=\tilde{n}+\delta\tilde{f}$, where 
$\tilde{n}(p,x)$ and $n(p,x)$ correspond to the thermal distributions of fermions and bosons
in the vicinity of the rest frame. 
We then assume that the perturbation takes the form,
\begin{eqnarray}
\delta \tilde{f}(p,x)=(1-\tilde{n}(p,x))\tilde{n}(p,x)\chi(p,x), 
\end{eqnarray}
where
\begin{eqnarray}\label{chi_decomp}
\chi(p,x)=\frac{B(p)}{T}\hat{p}^i\hat{p}^j\partial_iu_j,
\end{eqnarray}
with $u_j$ being the velocity and $T$ being the temperature of the medium.
Here $B(p)$ is associated with $\eta$ and the part corresponding to the bulk viscosity is dropped by setting $\nabla\cdot {\bf u}=0$ for simplicity. Collecting the leading-order contributions in (\ref{Boltzmann_eq}), we find
\begin{eqnarray}\nonumber\label{Boltzmann2}
\left(\frac{\delta^{ij}|{\bf p}|^2}{3}-p^ip^j\right)&=&e^2\int\frac{d^3{\bf q}\,\,\epsilon(q_0)}{(2\pi)^3 E_{p-q}}\bigg[\frac{\text{Im}\left[\Pi^R_{\mu\nu}(q)\right]}{(q^2)^2}\mathcal{P}^{\mu\nu}
\\
&&\times \frac{n(q,x)\tilde{n}(p-q,x)}{\tilde{n}(p,x)}\mathcal{B}_{ij}(p,q)\bigg],
\end{eqnarray}
where 
\begin{eqnarray}\nonumber
\mathcal{P}^{\mu\nu}&=&(2p^{\mu}(p-q)^{\nu}-p\cdot(p-q)g^{\mu\nu}+m^2g^{\mu\nu}),
\\\nonumber
\mathcal{B}_{ij}(p,q)&=&\left(B_{ij}(p)-B_{ij}(p-q)\right),
\\
B_{ij}(p)&=& B(p)\left(\hat{p}^i\hat{p}^j-\frac{\delta^{ij}}{3}\right).
\end{eqnarray}
From the definition of $\eta$ as the perturbation of the energy-stress tensor, one obtains
\begin{eqnarray}\label{shearviscosity}
\eta=\frac{\beta}{15}\int\frac{d^3{\bf p}}{(2\pi)^32E_p}|{\bf p}|^2(1-\tilde{n}_p)\tilde{n}_pB(p).
\end{eqnarray}

The primary task now is to solve the integral equation (\ref{Boltzmann2}) for $B(p)$. In general, the integral equation must be solved numerically \cite{Arnold:2000dr,Arnold:2003zc}. Nonetheless, in order to make manifest comparison with the result of the QED plasma, we may choose an appropriate ansatz for $B(p)$ and work out an approximate analytic solution of $\eta_{\text{mix}}$ up to the leading logarithmic order. 

From the structure of the integrand in the momentum integral of (\ref{Boltzmann2}), one may anticipate the dominance of low momentum transfer. This is true for QED and weakly coupled QCD plasmas. It also turns out to be the case when the QCD sector is strongly coupled. Therefore, we will work in the long-wavelength region for light leptons ($m \ll T$),
\begin{eqnarray}
q_0<|{\bf q}|\ll T\leq |{\bf p}|.
\end{eqnarray}
For convenience, we may further decompose  $\Pi^R_{\mu\nu}(q)$ into the transverse and longitudinal parts,
\begin{eqnarray}
\Pi^R_{\mu\nu}(q)=\hat{P}_{\mu\nu}^T\Pi^T(q)+\hat{P}_{\mu\nu}^L\Pi^L(q),
\end{eqnarray}
where $\hat{P}_{\mu\nu}^{T(L)}$ are projection operators. In the long-wavelength limit, (\ref{Boltzmann2}) reduces to
\begin{eqnarray}\label{Beq3}\nonumber
\left(\frac{\delta_{ij}|{\bf p}|^2}{3}-p_ip_j\right)&\approx& e^2\int\frac{d|{\bf q}|ds}{2\pi^2|{\bf q}|^3s}\bigg[\frac{T|{\bf p}|(1-s^2)\epsilon(q_0)}{\left((s^2-1)-\frac{m^2}{2|{\bf p}|^2}\right)^2}
\\
&&\mathcal{B}_{ij}(p,q)\text{Im}\left[\Pi^T(q)-\Pi^L(q)\right]
\bigg],
\end{eqnarray}
where we preserve the small lepton mass in $q^2$ to avoid the collinear divergence at $s=\cos\theta=\pm 1$ in the photon propagators
\footnote{Instead of introducing the
lepton mass, one could also employ the thermal-photon propagator to tackle the collinear divergence in the
$s$ integral, whereas the result remains unchanged up to the leading logarithmic order.}.

In order to study the effect of the non-perturbative dynamics in the QCD sector, we take the D3/D7-brane system in gauge/gravity duality as a concrete example \cite{Mateos:2006nu}, which is dual to a strongly coupled supersymmetric gauge plasma with quarks in the fundamental representation. Such a system as an analogue of the strongly coupled QGP in holography was frequently used to study electromagnetic signatures emitted from a thermal medium \cite{Mateos:2007yp,Wu:2013qja,Muller:2013ila}. In the long-wavelength limit for $q=(q_0,0,0,q_z)$, one finds \cite{Policastro:2002se}
\begin{eqnarray}\label{Pisub}
\text{Im}\left[\Pi^T(q)-\Pi^L(q)\right]
\approx-\frac{Q^2N_cN_fT}{4\pi}\frac{q_z^2\,\,\epsilon(q_0)}{q_0}
\end{eqnarray}
where the extra factor $\epsilon(q_0)$ is introduced for the cut correlators 
\footnote{In comparison with the D3/D7 setup in \cite{Mateos:2007yp}, one should make a replacement for the prefactors of the result in \cite{Policastro:2002se} by making $N_c^2\rightarrow 4Q^2N_cN_f$.}. 
Now, we may choose the simple {\em ansatz}
\begin{eqnarray}\label{Bansatz}
B(p)=C\frac{|{\bf p}|^2}{T^2\ln\left(\frac{2|{\bf p}|}{m}\right)},
\end{eqnarray}
where $C=2\pi^3/(e^2Q^2N_cN_f)$ is a dimensionless constant, which serves as the leading-order solution of (\ref{Boltzmann2}) in the long-wavelength limit. 

$B(p)$ diverges in the limit $m\to 0$ owing to the collinear divergence in the integral over $s$. As a regularization we may take $m\sim eT$ as the thermal mass of leptons, which is equivalent to introducing an IR cutoff at $|{\bf q}|\sim eT$ to regularize the IR divergence in QED. We could further take $\ln(2|{\bf p}|/(eT))\approx \ln(1/e)$ as an approximation applied in QED up to the leading logarithmic order. There exists a caveat that such an approximation may break down when $|{\bf p}|\geq T/e$. For quantitative correctness, one may also have to incorporate the Landau-Pomeranchuk-Migdal (LPM) effect with multiple Coulomb scattering for high-energy leptons. 
    
Inserting (\ref{Bansatz}) into (\ref{shearviscosity}), we obtain the $\eta_{\text{mix}}$ in a strongly coupled supersymmetric plasma modeled by the D3/D7-brane system,
\begin{eqnarray}\label{eta_mix}
\eta^{\text{D3/D7}}_{\text{mix}}\approx\frac{\pi T^3I_p}{30N_cN_fe^2Q^2\ln(1/e)},
\label{etamix}
\end{eqnarray}
where $I_p\approx 116$. 
To make a comparison with $\eta_{\text{QED}}$ on equal footing, one could follow the same approach and utilize the photon self energy in $\mathcal{O}(e^2)$ to solve the Boltzmann equation with a similar ansatz, which gives rise to 
\footnote{Although the $\eta_{QED}$ here qualitatively agrees with the one in \cite{Arnold:2000dr} at the leading logarithmic order, the prefactors differ. Since we only introduce the fluctuations of the "probe" electrons but neglect the fluctuations of the scattered electrons, the magnitude of $\eta_{QED}$ here is underestimated compared to \cite{Arnold:2000dr}. In other words, we here actually compute the shear viscosity of an energetic electron probing a low-energy QED plasma.}
\begin{eqnarray}\label{eta_QED}
\eta_{\text{QED}}=\frac{T^3I_p}{5\pi e^4\ln(1/e)}.
\label{etaQED}
\end{eqnarray}
Comparing (\ref{etamix}) with (\ref{etaQED}) we conclude that the ratio $\eta^{\text{D3/D7}}_{\text{mix}}/\eta_{\text{QED}}$ is suppressed in the large-$N_c$ limit.

Based on the dominance of the collinear divergence in our derivation, we can generalize our result by connecting $\text{Im}[\Pi^T(q)]$ in the long-wavelength limit with the direct-current (DC) electrical conductivity $\sigma_c$ of the QCD sector. The collinear divergence occurs when $q_0\rightarrow |{\bf q}|$, which yields $\text{Im}[\Pi^L(q)]\rightarrow 0$. Therefore, the integral on the left hand side of (\ref{Beq3}) will be dominated by the contribution from $\text{Im}[\Pi^T(q)]$.
Recall that the DC conductivity could be defined as \cite{CaronHuot:2006te}
\begin{eqnarray}
\sigma_c=\frac{1}{4}\lim_{q_0\rightarrow 0}\frac{\epsilon(q_0)\chi(q_0,|{\bf q}|)_{|{\bf q}|=q_0}}{q_0},
\end{eqnarray}
where $\chi(q_0,|{\bf q}|)$ denotes the spectral density, which can be computed from the retarded correlator,
\begin{eqnarray}
\chi(q_0,|{\bf q}|)=-2\text{Im}\left(2\Pi^T(q)+\Pi^L(q)\right).
\end{eqnarray}
The static limit ($q_0\rightarrow 0$) is approximately equivalent to the long-wavelength limit here. 
By dropping $\text{Im}[\Pi^L(q)]$ and replacing $\text{Im}[\Pi^T(q)]$ with the DC conductivity as
\begin{eqnarray}
\sigma_c=-\left(\frac{\epsilon(q_0)\text{Im}[\Pi^T(q)]}{q_0}\right)_{q_0\rightarrow|{\bf q}|\ll T},
\end{eqnarray}
we thus obtain the more general result:
\begin{eqnarray}\label{eta_conductivity}
\eta_{\text{mix}}\approx\frac{T^4I_p}{120\sigma_ce^2\ln(1/e)}.
\end{eqnarray}
  
Armed with (\ref{eta_conductivity}), we can evaluate $\eta_{\text{mix}}$ in the leading logarithmic order in different systems provided $\sigma_c$ is known. 
In pQCD at finite temperature, the DC conductivity up to the leading logarithmic order 
\cite{Chen:2013tra,Jiang:2014ura}
leads to 
\begin{eqnarray}
\eta^{\text{pQCD}}_{\text{mix}}\approx \frac{T^3I_p(N_c^2-1)g^4\ln(1/g)}{c_0N_c^2N_fe^2Q^2\ln(1/e)},
\end{eqnarray}
where $c_0$ is a numerical constant\footnote{Recall that we assume the quarks with different flavors carry same charge $Q$. When different flavors correspond to different charges, $N_fQ^2$ is replaced by $\text{Tr}_fQ^2$, where $Q$ denotes the charge matrix in flavor space \cite{Jiang:2014ura}.}. 
By contrast, in the Sakai-Sugimoto (SS) model of holographic QCD \cite{Sakai:2004cn,Sakai:2005yt}, the known value of the DC conductivity in the deconfined phase \cite{Parnachev:2006ev,Pu:2014cwa,Pu:2014fva}
yields  
\begin{eqnarray}
\eta^{\text{SS}}_{\text{mix}}\approx\frac{9\pi I_pT^2M_{\text{KK}}}{40\lambda N_ce^2Q^2\ln(1/e)},
\end{eqnarray}
where $\lambda$ denotes the t'Hooft coupling and $M_{\text{KK}}$ corresponds to the Kaluza-Klein mass as the meson mass scale in the gauge theory. Qualitatively, the ratio $\eta_{\text{mix}}/\eta_{\text{QED}}$ in pQCD and that in the SS model despite the competition of $M_{\text{KK}}$ and $T$ are $g^4\ln(1/g)$ and $1/(\lambda_t N_c)$ suppressed, respectively. These results may be representative of the viscous behavior of thermal leptons in high-temperature (pQCD) and strongly coupled, intermediate-temperature regimes of QGP.
   
\begin{figure}[t]
{\includegraphics[width=5.5cm,height=3.5cm,clip]{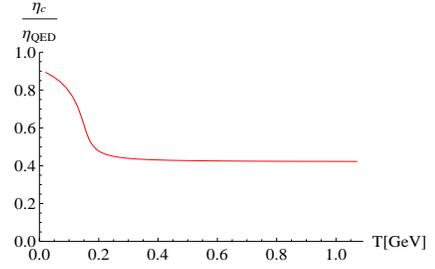}}
\caption{The ratio of the complete shear viscosity of leptons in a strongly coupled non-conformal plasma to that in the QED plasma versus temperature.}\label{viscosity_hQCD_fig}
\end{figure}

Using the analytical expressions derived above, we may attempt a crude estimate of the ratio $\eta_{\text{mix}}$ in the cosmological QGP beyond the deconfinement transition temperature. According to various approaches for the electrical conductivity of QGP \cite{Greif:2014oia}, one finds 
$\sigma_c/T\sim 0.02-0.1$ from $T=0.2$ GeV to $T=0.6$ GeV. This implies a ratio $\eta_{\text{mix}}/\eta_{\text{QED}}\sim0.12-0.6$, corresponding to $\eta_{c}/\eta_{\text{QED}}\sim0.11-0.38$. We conclude that the leptonic visosity in cosmological QGP was dominated by the interactions between leptons and quarks.
Nevertheless, $\sigma_c/T$ drops rapidly near the deconfinement transition, where the lepton-lepton scattering suppresses the lepton-quark interaction and $\eta_c/\eta_{\text{QED}}$ thus increases. As shown in Fig.\ref{viscosity_hQCD_fig}, by utilizing $\sigma_c/T$ obtained from the strongly coupled non-conformal gauge theory in holography \cite{Finazzo:2013efa}, $\eta_c/\eta_{\text{QED}}$ rapidly increases near the transition temperature. 

We now apply the same considerations to the interaction of hard leptons with the QGP. Because here we do not assume that the leptons are thermal, this calculation also has relevance to the QGP produced in relativistic heavy ion collisions.  For energetic or heavy leptons such as muons which slowly thermalize, we may compute the radiation rate $\Gamma^{>}(E_p)$ by simply replacing the thermal cut- propagator of leptons in the previous computations with the one in vacuum. Following the definition in \cite{Bellac:1996} with the long-wavelength approximation, the energy loss for a hard lepton traveling along the $z$ direction in the strongly coupled supersymmetric plasma characterized by the D3/D7 system could be written as
\begin{eqnarray}\label{Eloss_D3D7}\nonumber
\frac{dE_p}{dz}
&=&\frac{e^2Q^2N_cN_fT}{16\pi^3}\int d|{\bf q}|
ds\Bigg[\frac{s^2\left(1+D^2|{\bf q}|^2\right)}{v^2s^2+D^2|{\bf q}|^2}
\\
&&\left(\frac{v^3(1-s^2)}{(1-v^2s^2)^2}+\frac{v(1-v^2)}{(1-v^2s^2)}\right)\Bigg],
\end{eqnarray} 
where $s=\cos\theta$ and $v=|{\bf p}|/E_p$ and $D=(2\pi T)^{-1}$ is the diffusion constant. We focus on the large-velocity and small-velocity regions, where the analytic expressions are accessible.
For light leptons $m\sim eT\ll |{\bf p}|$, one could simply drop the diffusion terms and obtain  
\begin{eqnarray}\label{Eloss_D3D7_vmax}
\left(\frac{dE_p}{dz}\right)_{\text{D3/D7}}
\approx \frac{e^2Q^2\ln(1/e)N_cN_fT^2}{8\pi^3}\quad\text{for}\quad v\rightarrow 1.
\end{eqnarray}
For heavy leptons with $v\ll D|{\bf q}|$, the presence of the diffusion term prevents the $1/v$ divergence in (\ref{Eloss_D3D7}), while there still exists linear divergence in the IR regime for the $|{\bf q}|$ integral. By simply taking the IR cutoff $q_{IR}\sim eT$,
we acquire
\begin{eqnarray}
\left(\frac{dE_p}{dz}\right)_{\text{D3/D7}}\approx\frac{eQ^2N_cN_fT^3}{6\pi E_p}\quad\text{for}\quad v\rightarrow\frac{T}{E_p}.
\end{eqnarray} 
Compared with the energy loss in the leading order of two limits in the QED plasma,
\begin{eqnarray}\nonumber
\left(\frac{dE_p}{dz}\right)_{\text{QED}}&\approx& \frac{e^4\ln(1/e)T^2}{24\pi}\quad\text{for}\quad v\rightarrow 1,
\\
\left(\frac{dE_p}{dz}\right)_{\text{QED}}&\approx& \frac{e^4\ln(1/e)T^3}{36\pi E_p}\quad\text{for}\quad v\rightarrow\frac{T}{E_p},
\end{eqnarray}
it turns out that the pattern of the energy loss of light leptons in a QED plasma is similar to that in the strongly coupled supersymmetric plasma. Nonetheless, despite the prefactor altered by the choice of $q_{IR}$, the diffusion at small momentum transfer further enhances the energy loss for heavy leptons in the strongly coupled scenario. Even for light leptons, the energy loss is $\mathcal{O}(N_c)$ enhanced in the D3/D7 system, which is in accordance with the $1/N_c$ suppression of $\eta^{\text{D3/D7}}_{\text{mix}}/\eta_{\text{QED}}$. 

In summary, we have studied the shear viscosity and energy loss of leptons in strongly coupled (weakly coupled) QCD-like plasmas. It turns out that the non-perturbative effect of the QCD sector substantially affects the leptonic transport, rendering the lepton fluid more viscous.  In order to obtain more accurate results for the leptonic viscosity, numerical solvers of the Boltzmann equation beyond the leading logarithmic order are required. In addition, more realistic simulations of the current-current correlator should be involved.  Our approach can be easily generalized to other fluids that weakly interact with a strongly coupled medium. For instance, in the semi-holographic model of QGP \cite{Iancu:2014ava}, where hard and soft gluons are connected by effective perturbative couplings, our approach could be utilized to study the transport of hard gluons within the QGP. Moreover, there exist also thermal photons in the cosmic plasma, the study of the non-perturbative effect from the QCD sector on thermal-photon transport in the similar framework could be an interesting issue.   

{\em Acknowledgement}: The authors thank A. Mukhopadhyay for fruitful discussions and J. Noronha for providing the ratio of conductivity to temperature in the holographic model \cite{Finazzo:2013efa}. This work was
supported by Grant no. DE-FG02-05ER41367 from the
U. S. Department of Energy and in part by European Union's Seventh Framework Programme under grant agreements (FP7-REGPOT-2012-2013-1) no 316165, the EUGreece
program "Thales" MIS 375734 and was also co-financed by the European Union (European Social Fund, ESF) and Greek national funds through the Operational
Program "Education and Lifelong Learning" of the National Strategic Reference
Framework (NSRF) under "Funding of proposals that have received a positive
evaluation in the 3rd and 4th Call of ERC Grant Schemes".

\end{document}